\documentclass[twocolumn]{aastex631}
\usepackage{amsmath}
\usepackage{wasysym}
\usepackage{graphicx}
\usepackage{amssymb}
\usepackage{epstopdf}
\usepackage{mathrsfs}
\usepackage{anyfontsize}
\usepackage{natbib}
\usepackage{CJK}

\usepackage{color}
\usepackage{lipsum}
\usepackage{diagbox}
\usepackage{gensymb}
\usepackage{booktabs}
\usepackage{threeparttable}
\usepackage[figuresright]{rotating}
\usepackage{multirow}

\newcommand{\hamr}{h$\alpha$mr}
\newcommand{\lamr}{l$\alpha$mr}
\newcommand{\hamp}{h$\alpha$mp}
\newcommand{\lamp}{l$\alpha$mp}
\shorttitle{}
\shortauthors{Hu et al.}
\begin{document}
%\linenumbers
\begin{CJK*}{UTF8}{gbsn}
	\title{Quantifying chemical and kinematical properties of Galactic disks}
	\correspondingauthor{Zhengyi Shao}
	\email{zyshao@shao.ac.cn}
		
	\author[0000-0003-1828-5318]{Guozhen Hu (胡国真)}
	\affiliation{Key Laboratory for Research in Galaxies and Cosmology, Shanghai Astronomical Observatory, Chinese Academy of Sciences, 80 Nandan Road, Shanghai 200030, China}
	\affiliation{University of Chinese Academy of Sciences, 19A Yuquan Road, 100049, Beijing, China}
	
	\author[0000-0001-8611-2465]{Zhengyi Shao (邵正义)}
	\affiliation{Key Laboratory for Research in Galaxies and Cosmology, Shanghai Astronomical Observatory, Chinese Academy of Sciences, 80 Nandan Road, Shanghai 200030, China}
    \affiliation{Key Lab for Astrophysics, Shanghai 200234, China}

\begin{abstract}

 We aim to quantify the chemical and kinematical properties of the Galactic disks with a sample of 119,558 giant stars having abundances and 3D velocities taken or derived from the APOGEE DR17 and Gaia EDR3 catalogs. The Gaussian Mixture Model is employed to distinguish the high-$\alpha$ and low-$\alpha$ sequences along the metallicity by simutaneously using the chemical and kinematical data. Four disk components are identified and quantified that named as \hamp, \hamr, \lamp, and \lamr\ disks, which correspond to the features of  high-$\alpha$ or low-$\alpha$, and metal-poor or metal-rich. Combined with the spatial and stellar age information, we confirm that they are well interpreted in the two-infall formation model. The first infall of turbulent gas quickly forms the hot and thick \hamp\ disk  with  consequent thinner  \hamr\ and \lamr\ disks. Then the second gas accretion forms a thinner and outermost \lamp\ disk. We find that the inside-out and upside-down scenario does not only satisfy the overall Galactic disk formation of these two major episodes, but also presents in the formation sequence of three inner disks. Importantly, we reveal the inverse Age-[M/H] trend of the \lamr\ disk, which means its younger stars are more metal-poor, indicating that the rejuvenate gas from the second accretion gradually dominates the later star formation. Meanwhile, the recently formed stars convergence to [M/H]$\sim$-0.1 dex, demonstrating a sufficiently mixture of gas from two infalls.   
 
\end{abstract}
	

\section{Introduction}\label{sec:intro}

The Milky Way (MW) is a unique laboratory to study galaxy formation and evolution, where stellar properties can be measured and analyzed in great detail. Galactic archaeology utilizes stellar abundances and kinematics as relics to investigate the history of the MW, which relies on the assumption that any history of the Galactic formation and dynamical evolution will leave its marks on chemical abundances and kinematics, as well as their correlations, of stellar populations \citep{2002ARA&A..40..487F}.\\

In investigating the assembly history of the MW, it is essential to separate different components (i.e., stellar populations). The traditional view holds that there are two disk components, a thin disk and a thick disk in the MW. \cite{1983MNRAS.202.1025G} revealed this pattern about 40 years ago, and their work showed that at the Galactic South pole, the vertical stellar density profile could not be described by only one exponential. Then, the traditional definition of thin and thick disk features have been widely used: the thick disk is older than the thin disk (e.g., \citealt{1988AJ.....95.1404W,2013A&A...560A.109H,2014A&A...562A..71B,2019AJ....158...61B}); has a larger rotational lag compared to the thin disk (e.g., \citealt{2015A&A...583A..91G,2016A&A...586A..39R, 2020MNRAS.494.3880B, 2020A&A...643A.106B}); and has a higher velocity dispersion than the thin disk (e.g.,\citealt{2014A&A...567A...5R,2015A&A...583A..91G,2016MNRAS.461.4246W}). \\

Earlier studies used kinematics to separate thick and thin disk populations in two different ways. The first is to assume that the space velocities of different populations have different three-dimensional Gaussian distributions  \citep{2005A&A...438..139S,2006MNRAS.367.1181B,2010ApJ...721L..92R,2011A&A...525A..90K,2011A&A...535A.107K}. The second is to distinguish different components directly based on the Tommre diagram \citep{2019MNRAS.484L..69R}. However, as mentioned by previous studies (e.g., \citealt{2006MNRAS.367.1329R,2014A&A...562A..71B}), those distinctions based on kinematical criteria are imperfect. That is because the thin and thick disks overlap in their kinematical distributions, and the heating or migration processes might enhance such a mixture. Then, it might misidentify transitional stars, which perhaps bias conclusions \citep{2012ApJ...753..148B}. \\

With the accumulation of high-resolution spectroscopic data from large survey projects, studies show that the $\alpha$-enhancement ([$\alpha$/M]) and the metallicity ([M/H]) of local stars follow two distinct tracks (e.g., \citealt{2011A&A...535L..11A,2014A&A...562A..71B}), which provide much information about stellar populations, such as the star formation efficiency, the accretion history, and the gas outflows (e.g., \citealt{2017ApJ...835..224A,2017ApJ...837..183W}). The $\alpha$-elements are produced mainly during the Type II supernovae (SNe) explosions on a short time scale, whereas other elements (e.g., the iron) are mainly produced by Type Ia SNe on a much longer time scale. So the [$\alpha$/M] can be used as a 'clock' to probe the star formation history of a Galactic component \citep{2012A&A...538A..21S}. The high-$\alpha$ population is typically associated with the thick disk, while the low-$\alpha$ stars are associated with the thin disk. Moreover, many works realize that the dissecting method of components based on chemistry is superior to the method based on kinematics, since the chemical abundance is a relatively stable property of stars, which will not evolve after their formation. Thus, the mono-abundance populations (MAPs) are commonly used to trace the MW disk formation processes.  \\

At the metal-poor part ([M/H] $<-0.2$ dex), the stars are clearly separated in terms of [$\alpha$/M]. However, in the metal-rich range, the distinction is less obvious \citep{2004AN....325....3F,2011A&A...535L..11A}. \cite{2018ARA&A..56..223B} also revealed that the high-$\alpha$ and low-$\alpha$ sequences are merged up to [M/H] $\sim$ 0.15 dex. Some works suggested that the thick disk might not include stars with [M/H] $>-0.3$ dex \citep{2006MNRAS.367.1329R}. Usually, metal-rich stars with relatively higher [$\alpha$/M] are regarded as a particular population. Sometimes, the metal-rich range of the high-$\alpha$ sequence is called the 'transition zone'  (\citealt{2018MNRAS.476.5216D}, see their Fig. 8), or is classified as the so called \hamr\ disk (e.g. \citealt{2011A&A...535L..11A,2013A&A...554A..44A}, see their Fig. 1;  \citealt{2021ApJ...912..106Y}, see their Fig. 3), or is named as a 'bridge' component \citep{2021MNRAS.503.2814C}. In the metal-rich range, most previous works artificially distinguish high- or low-$\alpha$ components (thick or thin disks) with straight lines on [$\alpha$/M]-[M/H] plane (e.g., \citealt{2011ApJ...738..187L,2014A&A...564A.115A,2019MNRAS.484L..69R}). Definitely, such a simple cut may bias the properties of the real \hamr\ components and also contaminate the thin disk at the metal-rich part. Therefore, it is valuable to accurately determine the disks in the metal-rich region, if there are surely two separate populations. \\

The velocity and chemical abundance of stars are thought to involve clues to the formation and evolution of the MW. Stellar populations with different kinematics and abundances are related to different formation epochs. The origin of thick disk is still a matter of debate. A possible scenario is that the thick disk stars are formed by the dynamical heating of a pre-existing thin disk through small satellite mergers (e.g., \citealt{2008ApJ...688..254K,2009MNRAS.399.1145S,2021MNRAS.507.5882S}) or/and directly accreted from disrupted satellites (relatively metal-poor) around the MW (\citealt{2003ApJ...591..499A,2013ApJ...773...43B}). Another scenario is based on the two-infall model  (e.g., \citealt{1997ApJ...477..765C, 2017MNRAS.472.3637G,2019A&A...623A..60S,2020MNRAS.497.2371L,2020A&A...635A..58S}). In this model, the thick disk is due to an earlier gas infall which occurred a strong star burst and a subsequent secular star forming process. Then, a second infall rejuvenates the metal-poor gas to form the kinematically colder and thin disk. Anyway, for both of these scenarios, more detailed constraints should be based on the quantitative description of disks with their kinematical and chemical properties, and their coincidence. Therefore, it is necessary to investigate the separation and connection of disk components by simultaneously using kinematical and chemical data. \\

Some works have already combined the kinematical and chemical information in separating the thin and thick disks. \cite{2018MNRAS.476.5216D} dissected disks by constructing a mixture model in the space of [$\alpha$/M], [M/H] and radial velocity. They obscured the proper motion due to the  noteworthy observational errors at that time. In \cite{2021ApJ...921..106N}, they fitted a mixture of a number of Gaussian distributions using the three dimensional velocities and iron abundances of stars, while leaving the number of components as a free parameter. Unfortunately, these works have not adopted the full information of kinematics or chemical abundance. \\

In this work, we propose to quantitatively describe the disk chemical and kinematical properties with a sample of giant stars from the Apache Point Observatory Galactic Evolution Experiment (APOGEE) survey. Combining with the astrometric data from the Gaia, the sample stars have chemical measurements ([M/H] and [$\alpha$/M]) and three-dimensional velocity data. Then, for different [M/H] bins, we employ the Gaussian Mixture Model (GMM) to fit the star distribution in the remained four-dimensional data space, i.e., one for the [$\alpha$/M] and the other three for the velocities. This approach will allow us to curve the stellar population properties across the whole metallicity range and rigorously solve the mean values and their dispersions of these separated stellar components. It is worthwhile to mention that there is a potential advantage of using velocity and abundance distributions, since both of them are hardly affected by the incompleteness of the survey area, or the observational selection effects. \\

Additionally, based on the GMM framework, we can calculate the membership probabilities for individual stars of each component, i.e., belonging to the high-$\alpha$ or the low-$\alpha$ sequences. These membership probabilities are also valuable in further statistical analysis with other variables, e.g., the stellar age or their Galactocentric distances.\\

The paper is organized as follows. In Section 2, we describe the giant star sample used in this work, taken from the APOGEE and Gaia catalogs, supply with the derived distance, velocities, and stellar age measurements. Section 3 explains the strategy and method that we used to fit the data and assess the results. In Section 4, we present the fitting results leading to the classification of four disk components, the membership of individual stars, and discuss the effectiveness of our method. Then, We quantitatively describe the chemical abundance, kinematics, spatial and age distributions for each disk component and discuss the implications of the disk formation scenario in Section 5. Conclusions are summarized in Section 6. Throughout this work, we use [M/H] as the measure of metallicity and {[$\alpha$/M]} as a tracer of $\alpha$-enhancement.\\
		
\section{DATA AND SAMPLE SELECTION}\label{sec:samp}

Our sample is assembled by cross-matching APOGEE DR17 \citep{2021arXiv211202026A} and Gaia Early Data Release 3 (EDR3) \citep{2021A&A...649A...1G, 2021A&A...649A...2L}. The sample was then restricted to giant stars to avoid potential differential trends in the chemical measurements of dwarfs and giants. From the APOGEE data, we take the [M/H], [$\alpha$/M], and radial velocity of stars. The [M/H] and [$\alpha$/M] values are derived from the ASPCAP analysis of combined spectrum, and the $\alpha$-elements considered are O, Mg, Si, S, Ca and Ti. Combining with the coordination, Gaia proper motion, and distance \citep{2021AJ....161..147B} derived from the Gaia parallax, we calculate each star's position and velocity coordinates (the 6D phase space) and subsequently transfer them to the Galactocentric coordinate system. Moreover, we also collect stellar age measurements from different sources for further discussions.\\

\subsection{APOGEE DR17}\label{subsec:apogee}

APOGEE DR17 is a part of the final data release of the SDSS-IV (Sloan Digital Sky Survey; \citealt{2017AJ....154...28B}) via APOGEE-1 and APOGEE-2 programs. It contains 733,901 AllStar entries of 656,940 individual stars across both North and South hemispheres \citep{2021arXiv211202026A}. \\

Firstly, in order to avoid the bias of stellar element abundances obtained by ASPCAP when dealing with different types of stars \citep{2020ApJS..249....3A}, in this work, we only consider the giant stars ({\tt ASPCAP\_GRID} = g). Secondly, since we are only interested in the general structure of the MW and mainly focus on the disks, we exclude stars with  {\tt PROGRAMNAME} = bulge, gc, oc and halo-dSph. After that, the data quality is further controlled, and we only consider the stars with {\tt ASPCAP\_CHI2} $< 10$, {\tt ASPCAP\_FLAG} $= 0$, {\tt SNR} $> 70$, and {\tt M\_H\_ERR} $ < 0.1$ dex. Thus, we have totally picked up 235,250 giants from APOGEE DR17. For the duplicate observed sources, we use the error-weighted means instead.\\

\subsection{{Gaia} EDR3}\label{subsec:gaia}

{Gaia} EDR3 contains $\sim$1.81 billion sources with the magnitude completeness to $G \sim 20.7$ mag and has parallaxes and proper motions for a subset of $\sim$ 1.47 billion sources \citep{2021A&A...649A...1G, 2021A&A...649A...2L}. Gaia EDR3 extended the observational baseline to 34 months resulting in more precise astrometric measurements. Compared with the Gaia DR2, the parallax error of EDR3 was reduced by a factor of $\sim$20$\%$, and the accuracy of the proper motion was increased twice.\\

We cross-match the sample of giants from APOGEE DR17 with Gaia EDR3. We remain the stars with {\tt epsi} $<4$ and {\tt sepsi}  $<8$ \citep[][]{2021ApJ...908L...5S} to ensure the sources with acceptable astrometric solutions. In order to desert astrometric binaries and other anomalous sources, we retain only those stars with the re-normalized unit weight error {\tt RUWE} $<$ 1.4 \citep[][]{2021A&A...649A...2L}. \\

\begin{figure}[htbp]
	\centering
	\includegraphics[width=0.50\textwidth]{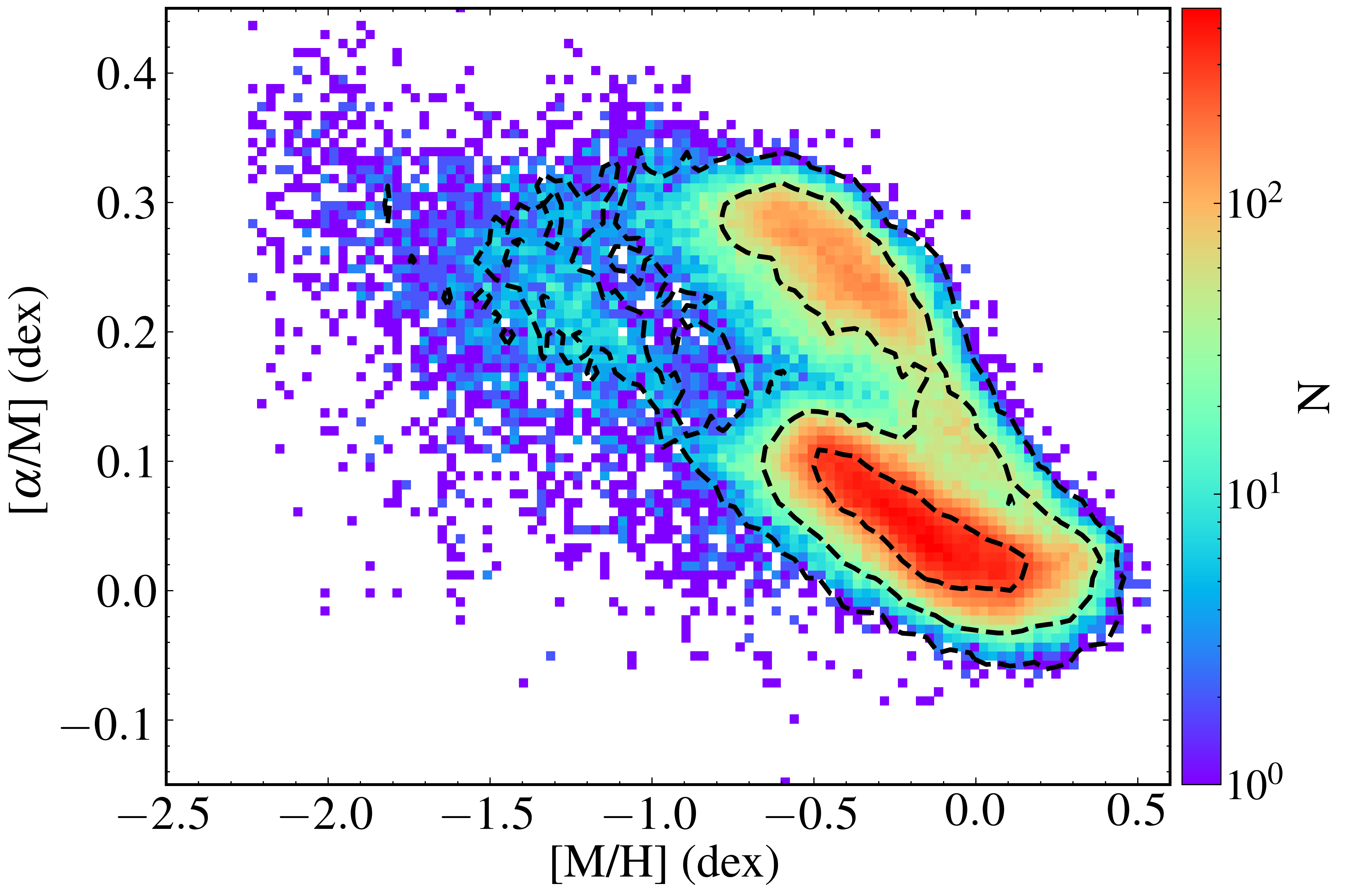}
	\caption{Number density distribution of sample stars in the [$\alpha$/M]-[M/H] plane. The colored distribution is for the full sample, and the dashed-lines correspond to the 1, 2 and 3$\sigma$ contours of stars with Age-S18 value.}

	\label{fig:all_sample_alphafeh}
\end{figure}

Finally, 119,558 stars are left in our sample. Fig.~\ref{fig:all_sample_alphafeh} shows the distribution of sample stars in the chemical abundance plane of  [$\alpha$/M] versus [M/H]. Generally, this diagram is applied to recognize the chemically defined thick (high-$\alpha$) and thin (low-$\alpha$) disk populations. The bimodal sequence of [$\alpha$/M] is very significantly shown by our sample. More interestingly,  there is a low number density area close to [M/H]$\simeq -0.1$ and [$\alpha$/M]$\simeq 0.14$, which implies a break of the star formation process. \\	

\subsection{Distance and velocities} \label{subsec:6D-position}

The coordination and proper motion of sample stars are directly taken from Gaia EDR3. We employ the photo-geometric distance estimated by \cite{2021AJ....161..147B} to avoid the parallax bias and high uncertainties of the Gaia targets far away from the Sun. Combined with the APOGEE's radial velocity, one can easily calculate the 6D kinematic information of a single star (3D positions and 3D velocities) related to the Sun. Then, we transform them to the Galactocentric cylindrical system $(R, \phi, Z, V_{R}, V_{\phi}, V_{Z})$ by assuming that the Galactocentric distance of the Sun is $R_\odot= 8.125$ kpc \citep{2018A&A...615L..15G} and its height from the Galactic plane is $Z_\odot=20.8$ pc \citep{2019MNRAS.482.1417B}. We also use a solar motion of (11.1, 242, 7.25) km\,s$^{-1}$ in the radial, rotational and vertical directions, respectively (\citealt{2010MNRAS.403.1829S,2019MNRAS.482.1417B}).\\

\subsection{Age measurements}\label{subsec:age}

The accurate stellar age determinations are decisive to investigate the MW formation and evolution \citep{2009ApJ...694.1498M,2010ARA&A..48..581S,2019MNRAS.487.3946M}. Unfortunately, stellar age is challenging to be measured directly (e.g., \citealt{1988AJ.....95.1404W,2013A&A...560A.109H,2014A&A...562A..71B}). It can only be estimated through model-dependent (e.g., \citealt{2005A&A...436..127J,2007ApJS..168..297T}) or empirical methods (e.g., \citealp{2007ApJ...669.1167B}).\\

In this work, we employ two recently published age catalogs. The first one we considered is from  \cite{2018MNRAS.481.4093S} (hereafter Age-S18). Based on the Bayesian framework, they used the photometric, spectroscopic, and astrometric information to characterize the stellar parameters. This catalog provides age estimation for $\sim$ 3 million stars. We reject stars with age uncertainties $\sigma_{\tau} >$ 1.5 Gyr. Then, 86137 APOGEE/Gaia/Age-S18 stars are left. In Fig.~\ref{fig:all_sample_alphafeh}, the dashed contour lines draw the distribution of the sample with Age-S18, which coincides with the colored map, indicating that the Age-S18 subset has no obvious selection bias.\\
	
The second age we used was taken from the SDSS DR17 Value-Added Catalog \citep{2019MNRAS.489..176M} (hereafter Age-M19). Age-M19 is estimated with a Bayesian neural network model \citep{2019MNRAS.489.2079L} trained by the asteroseismic ages \citep{2018ApJS..239...32P}. \cite{2019MNRAS.489..176M} claimed that the Age-M19 of stars with [Fe/H] $<$ -0.5 dex are less reliable than the others because of the lack of training set stars at the metal-poor region. The typical uncertainty of this age is 0.1 dex (25 percent). Totally, there are 118,766 APOGEE/{Gaia}/Age-M19 stars remaining in our sample.\\

\section{Methodology}\label{sec:method}

To model the distribution of our sample stars, we first divide them in [M/H] bins. Then, for each bin, we use the multivariate GMM to fit the subsample with four variables, [$\alpha$/M], $V_{R}, V_{\phi}$ and $V_{Z}$. This binning strategy is according to the following factors. Firstly, as Fig.~\ref{fig:all_sample_alphafeh} demonstrated, the [M/H] distributions are not regular (both high- and low-$\alpha$ sequences show 'banana' shapes), so they cannot be directly modeled as simple profiles. Secondly, [M/H] covers a wide range, and compared with [$\alpha$/M], has much smaller observational errors. Moreover, in the most part of the [$\alpha$/M]-[M/H] plane, for a given [M/H] value, the [$\alpha$/M] appears clearly bimodal shape. We suppose that each $\alpha$-enhancement sequence corresponds to a specific star formation process. Thus, for a given [M/H], the distribution of [$\alpha$/M] of each sequence is well expected to have a Gaussian profile (equal to the log-normal distribution of $\alpha/{\rm M}$), due to its complex origin through the star forming physical conditions.  Therefore, it will be very convenient to analyze the distribution structure with such binned subsamples.\\

\subsection{Gaussian Mixture Model}

A GMM is a parametric probability density function represented as a weighted sum of Gaussian components for multivariate densities. For the specific subsample within a given bin of [M/H], we assume that the stars follow a weighted sum of a number ($K$) of four dimensional (4D) normal distributions. If we further assume that this 4D normal distribution is independent of our adopting variables, [$\alpha$/M] and three velocities ($V_R, V_\phi, V_Z$), then it could be fully shaped with the parameters of mean values and dispersions for each one-dimensional normal distribution in each variable space, i.e., $ ({\boldsymbol{\mu}},{\boldsymbol {\sigma}}) = (\mu_{[\alpha/{\rm M]}}, \sigma_{[\alpha/{\rm M}]}; \mu_{V_R},\sigma_{V_R};\mu_{V_\phi},\sigma_{V_\phi};\mu_{V_Z},\sigma_{V_Z}) $.\\

For the $i$th star with observational data $\boldsymbol{x}_i = ([\alpha/{\rm M}]_i,V_{R,i},,V_{\phi,i},V_{Z,i})$, the likelihood of this star belonging to the GMM is written as:
%--
\begin{equation}
    \mathcal{L}_i = \sum_{k=1}^{K}  w_k \boldsymbol{\mathcal{N}}_k ( \boldsymbol{x}_i |\boldsymbol{\mu}_k,\boldsymbol {\sigma}_k) ,
\label{eq:Likelihood}
\end{equation}
%--
where the weights of each components satisfy
%--
\begin{equation}
    \sum_{k=1}^{K}  w_k  =1 .
\label{eq:GMM}
\end{equation}
%--
Then we can write down the joint logistical likelihood for the subsample with $N$ data points,
%--
\begin{equation}
    \log \mathcal{L} = \sum_{i=1}^{N} \log \mathcal{L}_i .
\label{eq:logLike}
\end{equation}

We employ the EM algorithm \citep{1986ApJ...306..490I} method to search the $\mathcal{L}_{\rm max}$ across all parameter ranges to obtain their optimal values. \\

It is clear that we need to somehow decide on the number ($K$) of Gaussian components by trading off quality of fit against the number of introduced free parameters. To do so, we use the Bayesian Information Criterion (BIC; \citealt{1978AnSta...6..461S,2000astro.ph..8187C}) to determine how many mixtures we should use. The BIC is defined as:
%--
\begin{equation}
    {\rm BIC} = - 2 \log \mathcal{L}_{\rm max} + m \log(N),
\label{eq:BIC}
\end{equation}
%--
where $m$ is the number of free parameters, and in this work, $m=3K-1$, for each sub-samples. Then, we compare the corresponding BICs with different numbers of components and select the model with the smallest BIC value. \\

\subsection{Membership probability and effectiveness}\label{subsec:member_eff}

For a set of optimal parameters of $(w_k,\boldsymbol{\mu}_j,\boldsymbol{\sigma}_k; k=1\,{\rm to}\,K)$,  the probability that a data point ($\boldsymbol{x}_i$ for the $i$th star) belonging to the $j$th population is:
%--
\begin{equation}
	P_j(i)= \frac{w_j \boldsymbol{\mathcal{N}}_j (\boldsymbol{x}_i |\boldsymbol{\mu}_j,\boldsymbol{\sigma}_j)} {\sum_{k=1}^{K} w_k \boldsymbol{\mathcal{N}}_k (\boldsymbol{x}_i |\boldsymbol{\mu}_k,\boldsymbol {\sigma}_k)}. \\
\label{eq:Pmemb}
\end{equation}
%--
Alternatively, one can calculate the $P_j$ value in the case of involving only a part of observational variables. For example, if we neglect the star's velocities and consider the [$\alpha$/M] data only, we can determine the GMM distributions in the [$\alpha$/M] space and calculate the corresponding probability  $P_j^{\alpha}$. Similarly, one can also calculate probability $P_j^{\boldsymbol{v}}$ as the probability that only using the velocity data. These different definitions of $P_j$ will be used to evaluate the improvement of the combination of multiple observational data (see Sec.~\ref{subsec:Eff-Result} for some examples). \\

It is clear that $P_j$ is a kind of normalized values through $K$ components. Based on the distribution of $P_j$ for the $j$th component, we can quantitatively evaluate how effective the results of our membership determination are \citep{1996AcASn..37..377S}. The effectiveness of membership determination is defined as:
%--
\begin{equation}
	E_j = 1-\frac{N\sum_{i=1}^N{P_j(i)[1-P_j(i)]}}	 {\sum_{i=1}^NP_j(i)\sum_{i=1}^N[1-P_j(i)]}.
\label{equ:eff}
\end{equation}
%--
The range of $E$ value is $[0,1]$. The bigger $E$ is, the more effective the membership determination is. In other words, $E$ could be regarded as an index to assess the mixture level of the GMM, e.g., a more discrete Gaussian component will lead to a bigger $E$ value from its membership probability distribution.\\

\section{Identification of Stellar Components and their fitting results}\label{sec:result}

\subsection{Number of mixed components}\label{subsec:number}

All of our sample stars have been binned in 0.1 dex interval of [M/H]. Totally, there are 28 bins from -2.3 to 0.5 dex of [M/H]. In this work, we need to choose one or two components to model the data within a given bin. So, we employ either one or two components GMM to fit each subsample and calculate the corresponding values of BIC$_1$ or BIC$_2$, and  make the choice of $K$ based on $\Delta {\rm BIC} = ({\rm BIC}_1 -{\rm BIC}_2)/N$. The $\Delta$BIC results are listed in table~\ref{tab:result}. Among them, there are 16 bins having $\Delta {\rm BIC} > 0$, all within the range from [M/H]=$-1.2$ to 0.4 dex. So, for these bins, the two Gaussian distributions ($K=2$) are adopted. According to the mean values of [$\alpha$/M], they are named as high- or low-$\alpha$ populations. For the other bins, at the metal-poor region ([M/H]$ < -1.2$ dex) and the most metal-rich bin ([M/H]$ \simeq 0.45$ dex), their $\Delta {\rm BIC} < 0$, which suggests it is unnecessary to separate the subsample into two populations.  \\

It should be mentioned that, at the metal-rich part ([M/H]$ > -0.1$ dex), although the bimodal feature of [$\alpha$/M] is not as clear as those at the more metal-poor region, the BIC values indicate that only one Gaussian is not enough to curve the distributions for most of these bins. Obviously, this is an evidence of none-single population from the data themselves. It implies the complicated composition of these metal-rich stars. In other words, the so-called \hamr\ disk might exist, but definitely could not be simply distinguished by straight lines on the [$\alpha$/M]-[M/H] plane. \\

\subsection{Classification and parameters of disk components}\label{subsec:Classification}

\begin{figure*}%[htbp]
	\centering
	\includegraphics[width=1.00\textwidth]{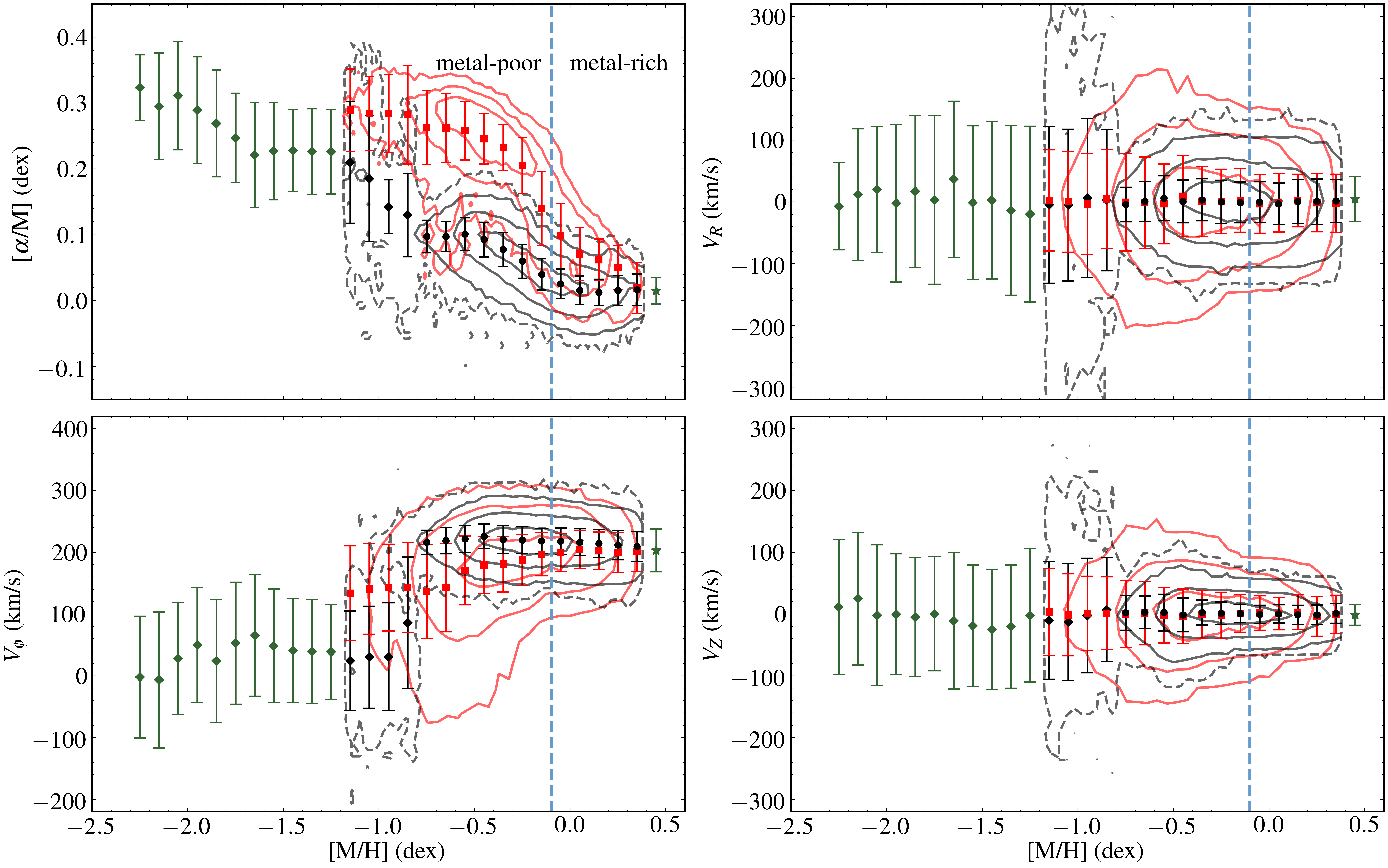}
	\caption{Results of the Gaussian mixture decomposition. The top-left panel is the projection on the [$\alpha$/M]-[M/H] plane. Other three panels are projections on three velocity-[M/H] planes respectively. The symbols with error bars show the GMM fitting results of the mean and dispersion values for each Gaussian component, with 'red' and 'black' for the bi-normal model ($K=2$), while 'green' for the single Gaussian model ($K=1$). Different symbols represent different populations, with diamonds for the halo, squares and circles for the high- and low-$\alpha$ sequences respectively, and the only green star for the most metal-rich sub-sample.  Contours show the probability-weighted number densities of high-$\alpha$ (red) and low-$\alpha$ (black) stars. The solid lines correspond the 1, 2 and 3$\sigma$ contours, and the black dashed line is the 4$\sigma$ contour of the low-$\alpha$ sequence to demonstrate the different distribution features  of halo and disk stars. The blue dashed vertical lines at [M/H]=-0.1 divide the metal-poor and metal-rich parts. 
}

	\label{fig:FittingResults}
\end{figure*}

The fitting results of parameters $(\boldsymbol{\mu},\boldsymbol{\sigma})$ are summarized in table~\ref{tab:result}. The mean and dispersion values are also shown as the symbols with error bars in Fig.~\ref{fig:FittingResults}. Since there are higher fitting precision, i.e., the uncertainties of all parameters are significantly smaller than the corresponding Gaussian dispersions, we omit the errors of fitting parameters either in the table or in the figure. \\

Based on these chemical and velocity parameters, we can classify each Gaussian component of the MW populations. For the metal-poor subsamples with [M/H]$<-1.2$ dex, $K=1$ for each bins. They have relatively higher [$\alpha$/M], roughly zero means and large dispersions ($\sim 100$ km s$^{-1}$) of all velocity components, $V_R$, $V_\phi$ and $V_Z$. Obviously, they are halo stars. In the range of $-1.2 < $[M/H]$ < 0.4$, where we should use two Gaussian components to model the subsamples, it can be further discriminated into several parts. For the high-$\alpha$ sequence, it could be split at [M/H]$ \sim -0.1$ dex. The metal-poor part of this sequence is regarded as the canonical thick disk, and we use \hamp\ to denote this component. For the metal-rich part of this sequence, the [$\alpha$/M] further decreases with metallicity, but still higher than another component in the same [M/H] bin. As noted by previous works (\citealt{2011A&A...535L..11A,2013A&A...554A..44A,2021ApJ...912..106Y}), we define this component as the \hamr\ disk. \\

The low-$\alpha$ sequence also has multiple compositions. The first split point is at [M/H]$\simeq -0.8$, where there is a clear break of the kinematic features, either in the mean value of $V_\phi$, or in all three velocity dispersions. The more metal-poor bins follow the kinematics of the halo component, though their [$\alpha$/M] values are lower than the corresponding thick disk stars. This phenomenon could be explained as the MW halo is a complicated population that contains quite a lot of accreted dwarf galaxies \citep{2010A&A...511L..10N}. The other more metal-rich bins of this sequence are similar, with their characteristics representing the canonical thin disk, either in chemical abundance or in kinematics. However, here we prefer to separate these bins into two groups by [M/H]$ \sim -0.1$.  The component with $-0.8 < $[M/H]$ < -0.1$ of the low-$\alpha$ sequence is regarded as the \lamp\ disk, while the more metal-rich component ([M/H]$ >- 0.1$) is named as the \lamr\ disk. According to the fitting parameters, there are slight differences between these two low-$\alpha$ disks. The \lamp\ disk shows a correction between [$\alpha$/M] and [M/H], with the more metal-rich stars having lower [$\alpha$/M] values. Whereas the [$\alpha$/M] values of the \lamr\ disk do not change with the metallicity. On the other hand, the $\sigma_{V_Z}$ of the \lamr\ disk is smaller than the \lamp\ disk. Rigorously speaking, there are overlaps between two low-$\alpha$ disks. However, considering the break in the [$\alpha$/M]-[M/H] correlation and other evidences from the age and spatial distributions (see detailed comparison and discussions in Sec.~\ref{sec:discussion} ), we use a simple cut at [M/H]=-0.1 to separate these two populations. \\

For the most metal-rich bin of our sample ([M/H]$\sim$0.45), the optimal distribution model is a single Gaussian. We suppose it is a merge of the \hamr\ and \lamr\ populations. \\

\begin{table*}[h] 
\setlength\tabcolsep{4pt}
	\setlength{\abovecaptionskip}{0.05cm}
	\centering
	\scriptsize
	\caption{The fitting results and mean ages in different metallicity bins} \label{tab:result}

	\begin{threeparttable}
		\begin{tabular*}{\hsize}{@{\extracolsep{\fill}} ccrcclrrrrrrrrcc} 
			\hline
			\hline
			[M/H] bin&N & $\Delta$BIC & K & $f_{h\alpha}$ & Comp.~\tnote{a} & [$\alpha$/M] & $\sigma_{[\alpha/M]}$ & $V_{R}$ & $\sigma_{V_{R}}$ & $V_{\phi}$ & $\sigma_{V_{\phi}}$ &  $V_{Z}$ & $\sigma_{V_{Z}}$ &Age-S18~\tnote{b}&Age-M19~\tnote{c}\\
			dex & -& -& -&-& -&dex& dex & ${\rm km\,s^{-1}}$ & ${\rm km\,s^{-1}}$& ${\rm km\,s^{-1}}$ &${\rm km\,s^{-1}}$ &${\rm km\,s^{-1}}$ & ${\rm km\,s^{-1}}$&Gyr&Gyr\\
			
			\hline

			[-2.3, -2.2] & 15 & -0.73 & 1& ... & halo& 0.323 & 0.050 & -7.2 & 70.6 & -2.0 & 98.7 & 11.4 & 109.7&10.85&8.85\\

			[-2.2, -2.1] & 41 & -0.30 & 1& ... & halo& 0.295 & 0.081 & 11.6 & 106.5 & -6.7 & 110.0 & 24.9 & 107.5&11.03&8.74\\

			[-2.1, -2.0] & 63 & -0.27 & 1& ... & halo & 0.311 & 0.082 & 20.0 & 102.3 & 27.9 & 90.7 & -1.8 & 113.7&11.10&8.86\\

			[-2.0, -1.9] & 89 & -0.32 & 1& ... & halo & 0.289 & 0.081 & -2.1 & 127.5 & 50.1 & 93.0 & -0.3 & 97.8&10.80&8.83\\

			[-1.9, -1.8] & 116 & -0.33 & 1& ... & halo& 0.269 & 0.081 & 16.9 & 122.9 & 24.4 & 99.5 & -5.2 & 96.4&10.84&8.53\\

			[-1.8, -1.7] & 140 & -0.24& 1 & ... & halo& 0.247 & 0.068 & 3.4 & 136.6 & 53.0 & 98.8 & 0.4 & 92.4&10.80&8.46\\

			[-1.7, -1.6] & 179 & -0.08& 1 & ... & halo & 0.221 & 0.080 & 36.5 & 126.6 & 65.4 & 98.3 & -10.7 & 110.6&10.28&7.98\\

			[-1.6, -1.5] & 263 & -0.12& 1 & ... & halo& 0.227 & 0.074 & -1.4 & 124.4 & 48.6 & 92.2 & -18.9 & 98.3&10.40&7.84\\

			[-1.5, -1.4] & 294 & -0.13& 1 & ... & halo & 0.228 & 0.062 & 2.6 & 127.4 & 41.2 & 84.3 & -25.0 & 97.0&9.05&7.51\\

			[-1.4, -1.3] & 359 & -0.10 & 1& ... & halo & 0.226 & 0.065 & -13.8 & 137.1 & 39.0 & 84.1 & -20.3 & 99.7&8.86&6.98\\

			[-1.3, -1.2] & 441 & -0.01 & 1& ... & halo & 0.226 & 0.064 & -19.8 & 142.4 & 38.8 & 76.8 & -2.0 & 107.8&8.40&6.24\\

			\hline

			[-1.2, -1.1] & 418 &0.08&2&	0.32& h$\alpha$mp&0.289& 0.062 &2.4 &82.0 & 133.9& 76.4 & 3.4 & 70.8&8.83&7.20\\
			~&~&~&~&~&halo&0.210&0.092&-4.8&126.8&24.6&80.4&-10.2&95.2&8.90&5.82\\

			[-1.1, -1.0] &
			422&
			0.02&
			2&
			0.38&

			h$\alpha$mp&0.284& 0.056 &0.2 &81.8 & 140.7& 73.2 & -1.3 & 66.3&8.63&6.90\\
			~&~&~&~&~&halo&0.185&0.095&-5.2&122.9&30.4&82.5&-13.0&94.8&8.85&5.96\\
			
			[-1.0, -0.9] &
			502&
			0.18&
			2&
			0.52&

			h$\alpha$mp&0.284& 0.059 &-3.5 &81.9 & 142.9& 70.2 & 0.9& 60.3&8.49&7.68\\
			~&~&~&~&~&halo&0.143&0.041&6.3&128.8&31.0&87.4&-2.3&92.8&8.07&5.37\\

			[-0.9, -0.8] &
			883&
			0.57&
			2&
			0.74&

			h$\alpha$mp&0.282& 0.074 &4.6 &80.3& 142.9& 73.0& 1.4& 59.0&8.17&7.51\\
			~&~&~&~&~&halo&0.130&0.063 &2.7&114.4&85.9&106.4& 7.0&84.2&6.66&4.37\\

			[-0.8, -0.7] &
			2040&
			0.64&
			2&
			0.86&

			h$\alpha$mp&0.263& 0.056&-0.5&79.0& 136.4& 76.1 & -0.1& 54.1&8.10&7.81\\
			~&~&~&~&~&l$\alpha$mp&0.098&0.025&-4.4&28.0&216.2&19.6&2.2&28.0&5.93&4.82\\

			[-0.7, -0.6] &
			4223&
			1.04&
			2&
			0.79&

			h$\alpha$mp&0.262& 0.052&-1.4 &73.9& 142.9& 71.7 & 1.7 & 51.5&7.79&8.16\\
			~&~&~&~&~&l$\alpha$mp&0.097&0.023&0.6 &32.3 &218.7 &21.2 &3.1&26.2&5.33&5.07\\

			[-0.6, -0.5] &
			8485&
			1.59&
	    		2&
			0.50&

			h$\alpha$mp&0.258& 0.045 &-3.2 &64.7 & 170.6& 55.5 & -2.0& 44.7&7.46&8.30\\
			~&~&~&~&~&l$\alpha$mp&0.101&0.025&5.7&36.4&221.7&21.6&2.5&29.8&5.30&5.01\\

			[-0.5, -0.4] &
			14110&
			1.62&
			2&
			0.36&

			h$\alpha$mp&0.246& 0.038 &9.3&65.7 & 179.3& 45.7 & -3.6& 44.4&7.07&8.42\\
			~&~&~&~&~&l$\alpha$mp&0.093&0.026&0.6&34.8&225.8&19.9&-1.5&27.3&4.83&4.47\\

			[-0.4, -0.3] &
			15871&
			1.48&
			2&
			0.31&

			h$\alpha$mp&0.232& 0.034&0.7 &58.9 & 180.9& 45.0 & -0.9 & 39.1&6.89&8.46\\
			~&~&~&~&~&h$\alpha$mp&0.078&0.026& 3.2 &34.0 &220.6&22.3&2.4&20.2&4.39&3.78\\

			[-0.3, -0.2] &
			16683&
			1.20&
			2&
			0.27&

			h$\alpha$mp&0.205& 0.040 &0.7&52.3 & 187.3& 41.2 & -1.7 & 37.3 &6.86&8.21\\
			~&~&~&~&~&l$\alpha$mp&0.060&0.026&-0.0&35.1&219.6&21.6&1.2&18.1&4.03&3.40\\

			[-0.2, -0.1] &
			16336&
			0.91&
			2&
			0.25&
			
			h$\alpha$mp&0.140& 0.056 &2.4 &51.3 & 196.6& 35.0 & 1.2 & 30.5 &6.37&7.44\\
			~&~&~&~&~&l$\alpha$mp&0.040&0.023&-1.0&33.1&218.3&20.6 &1.0&15.2&3.54&3.08\\

			[-0.1,$\hspace{0.35em}$ 0.0] &
			14137&
			0.59&
			2&
			0.25&
			
			h$\alpha$mr&0.098& 0.050 &-3.2&47.5& 199.7& 30.5 & 1.8& 27.8&6.14&7.12\\
			~&~&~&~&~&l$\alpha$mr&0.026&0.022&-0.7&33.0 &217.8&21.5&-0.4&14.9&3.57&3.41\\

			[0.0,$\hspace{0.7em}$ 0.1] &
			11257&
			0.36&
			2&
			0.27&
			
			h$\alpha$mr&0.071& 0.040 &-1.4 &47.3 & 204.7& 30.7 & 0.8 & 26.2 &5.82&6.64\\
			~&~&~&~&~&l$\alpha$mr&0.016&0.022&-3.3&33.7&216.3&21.6&0.3&14.8&3.59&4.08\\

			[0.1,$\hspace{0.7em}$ 0.2]&
			7270&
			0.21&
			2&
			0.18&
			
			h$\alpha$mr&0.062 & 0.031 & 2.2 &42.6 & 202.4 & 30.3  & 3.0 & 28.1 &5.63&6.60\\
			~&~&~&~&~&l$\alpha$mr&0.013&0.019&0.3&35.6&214.2&23.1&-1.1&15.5&4.08&4.65\\

			[0.2,$\hspace{0.8em}$ 0.3] &
			3490&
			0.15&
			2&
			0.17&

			h$\alpha$mr&0.050 & 0.033 &-1.4 & 47.2 & 199.4 & 32.6 & -3.6 & 32.2 &5.44&6.33\\
			~&~&~&~&~&l$\alpha$mr&0.016&0.023&0.4&36.7&211.6 &23.9&-1.3&16.0&4.57&5.35\\

			[0.3,$\hspace{0.7em}$ 0.4] &
			1310&
			0.38&
			2&
			0.08&

			h$\alpha$mr& 0.019 & 0.038 &-2.2  &46.9 & 201.3 & 32.0 & -0.5 & 25.1&5.07&5.78\\
			~&~&~&~&~&l$\alpha$mr&0.017&0.024&1.4&34.9&209.2&23.9&1.2&16.0&5.08&5.78\\

			\hline
			[0.4,$\hspace{0.8em}$ 0.5]&121&-0.38& 1& ... & mr & 0.015 &0.020& 4.5 & 36.7&203.0 &34.8& -1.4& 16.5&4.81&5.90\\
			
			\hline
			\hline
			
		\end{tabular*}
		\begin{tablenotes}
			\footnotesize
			\item[a] Classification of the MW components.
			\item[b] \citet{2018MNRAS.481.4093S}. The Age-S18 is probability-weighted average age of each component in each bin.
			\item[c] \citet{2019MNRAS.489..176M}. The Age-M19 is probability-weighted average age of each component in each bin. The Age-M19 of stars with [M/H]$< -0.5 $ dex are less reliable. 
		\end{tablenotes}	
	\end{threeparttable}
\end{table*}

\subsection{Membership probability of each component}\label{subsec:Pmemb}

Using Eq.~\ref{eq:Pmemb}, for the bimodal bins, we calculate the probability of each star belonging to the high-$\alpha$ sequence, $P_{h\alpha}$. Accordingly, the low-$\alpha$ probability, $P_{l\alpha} = 1-P_{h\alpha} $. The probability values are listed in table~\ref{tab:Pmemb}. Based on the disk classification, we can obtain the probabilities of each stars belonging to a specific disk or halo components. By summing of the probabilities, in our giant star sample, the fraction of different components are, halo (2.5\%), \hamp\ disk (24.7\%), \hamr\ disk (7.3\%), \lamp\ disk (41.4\%)  and \lamr\ disk (24.1\%).\\

\begin{table*}[h] 
\setlength{\abovecaptionskip}{0.05cm} 
\centering
\caption{Chemical and kinamatical data and membership probability of sample stars} \label{tab:Pmemb}

\begin{threeparttable}
\begin{tabular*}{\hsize}{@{\extracolsep{\fill}}llrrrrrrc} 
\hline
\hline

APOGEE-ID~\tnote{a}& SOURCE-ID~\tnote{b} & {[$\alpha$/M]}~\tnote{c}& {[M/H]}~\tnote{d} &  $V_{R}$ & $V_{\phi}$ & $V_{Z}$ & $K$ & $P_{h\alpha}$\\
 & & dex & dex & km\,s$^{-1}$ & km\,s$^{-1}$ & km\,s$^{-1}$ & & \\
\hline
2M10061549+4720367 & 822188376907896064 & 0.245 & -2.247 & -127.6 & 113.1 & -64.8 & 1 & ...\\
2M15505257+2756175 & 1223770877101987712 & 0.279 & -1.855 & 203.9 & 2.4 & 181.7 & 1 & ...\\
2M21024981-0118487 & 6917139903604771712 & 0.286 & -0.929 & -64.2 & -2.1 & -77.0 & 2 & 0.99\\
2M23472814+6149358 & 2012869825245864320 & 0.123 & -0.607 & 0.4 & 200.1 & -45.6 & 2 &0.02\\
2M02325239+2829337 & 128147809635150336 & 0.134 & -0.462 & -54.2 & 147.8 & -8.7 & 2 &0.90\\
2M07032260+5715085 & 1000544349938145920 & 0.109 & -0.205 & -98.6 & 198.0 & -24.0 & 2 &0.35\\
2M18461632-2259448 & 4078443430416791040 & 0.077 & -0.199 & -29.4 & 243.9 & 53.0 & 2 &0.82\\
2M14512752-5925456 & 5878295071153144192 & 0.041 & 0.140 & -42.5 & 225.9 & -9.8 & 2 &0.10\\
2M07060606-7440158 & 5261235195577098880 & 0.018 & 0.255& 45.8 & 214.3 & -40.7 & 2 &0.30\\
2M15181135-4920358 & 5902178284791783808 & 0.008 & 0.491 & 16.7 & 176.9 & 0.2 & 1 &...\\
\hline
\hline
\end{tabular*}
\tablecomments{The complete table is available on the publisher website.}
\begin{tablenotes}
	\footnotesize
	\item[a] APOGEE DR17 object name.
    \item[b] Gaia EDR3 source id.
	\item[c] [$\alpha$/M] in APOGEE DR17.
	\item[d] [M/H] in APOGEE DR17.	
\end{tablenotes}

\end{threeparttable}
\end{table*}

Figure~\ref{fig:FittingResults} also shows the probability weighted number densities of high- and low-$\alpha$ sequences in red or black contours. As expected, in the [$\alpha$/M]-[M/H] plane, these two sequences distinguish clearly in the region of more metal-poor than -0.1 dex, but still with unneglectable overlap. The overlapping phenomena are more significant in those $V$-[M/H] planes than in the [$\alpha$/M]-[M/H] plane. However, it shows differences in the velocity dispersions and the mean values of $V_\phi$. These kinematical differences enhance the distinctness of these two [$\alpha$/M] dominated sequences. In all, these two sequences have natural looking density profiles in either [$\alpha$/M]-[M/H] or $V$-[M/H] planes, which implies the combination of chemical and kinematic data is valuable. \\

Additionally, the probability values provide an approach to take into account all individual stars' contribution to a specific population. This illustrates the ability of mixture modeling to allow full probabilistic analysis, which is particularly significant when considering the degree of overlap between populations.\\

\subsection{Effectiveness of the combination of abundance and kinematics}\label{subsec:Eff-Result}

\begin{figure*}%[htbp]
	\centering
	\includegraphics[width=1.00\textwidth]{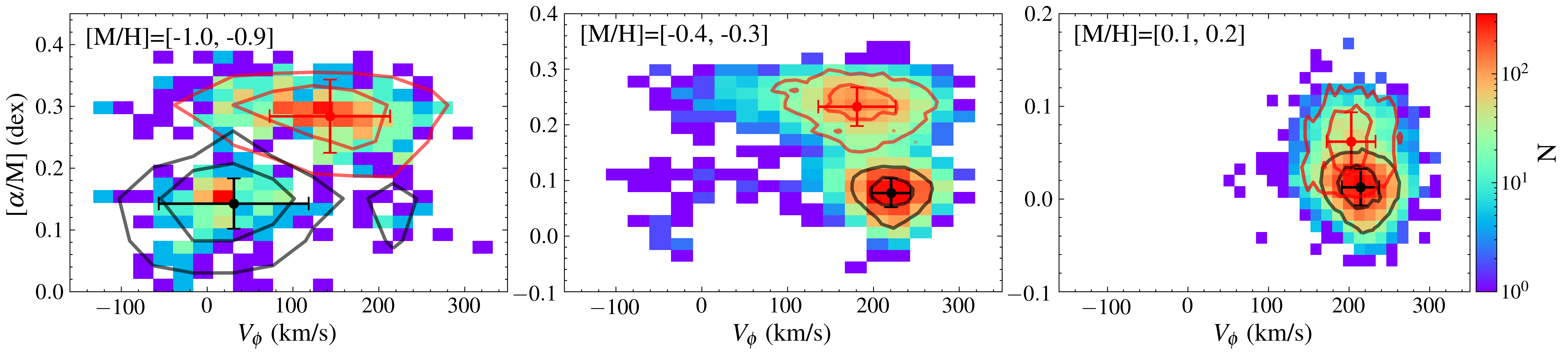}
	\caption{The number density and GMM fitting results for three representational metallicity bins ([-1, -0.9], [-0.4, -0.3], [0.1, 0.2]) in the $V_\phi$- [$\alpha$/M] plane. Red and black symbols with error bars show the mean and dispersion values of each Gaussian component. Red and Black lines are 1 and 2$\sigma$ contours of the probability-weighted number density for each component. } %{ bin fit. }
	\label{fig:binFit}
\end{figure*}

In Fig.~\ref{fig:binFit}, as examples, we plot the fitting results and number density distributions in the [$\alpha$/M]-$V_\phi$ plane of three representational subsamples ([M/H]$ \simeq -0.95$, -0.35 and 0.15 dex, for cases of the \hamp-halo, \hamp-\lamp\ and \hamr-\lamr\ mixtures), to demonstrate the reasonableness of using the bimodal distribution in the [$\alpha$/M] versus velocity space. Besides the most significantly discrete \hamp-\lamp\ distribution (in the middle panel), the other two subsamples are also shown indications of the bimodality in the [$\alpha$/M]-$V_\phi$ plane.\\

\begin{figure*}%[htbp]
	\centering
	\includegraphics[width=1.00\textwidth]{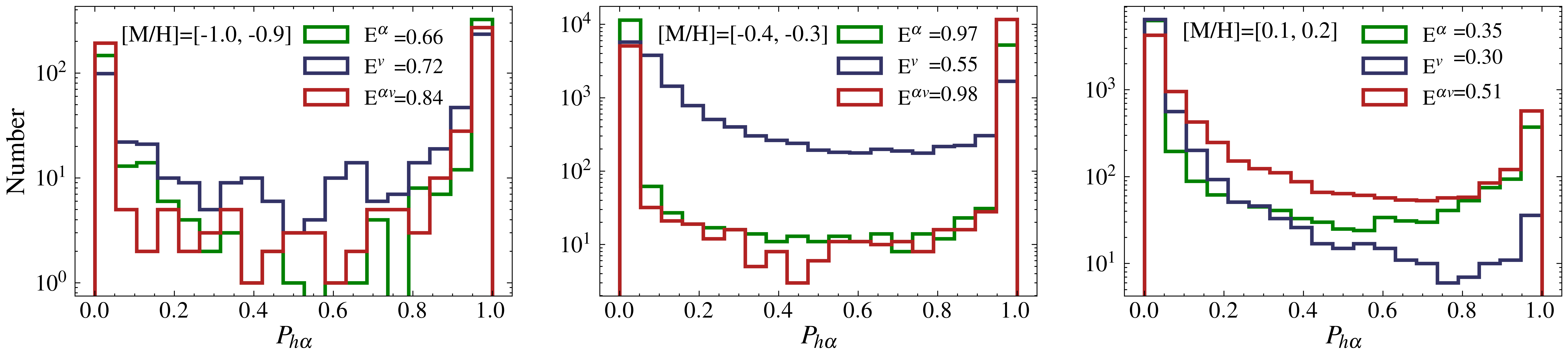}
	\caption{Probability distributions of the high-$\alpha$ components of three representational metallicity bins (same as in Fig.~\ref{fig:binFit}).  The green, blue and red histograms represent the membership probabilities using only velocity or chemical data, or both of them. The effectiveness indexes for each probability distribution are calculated.  }
	\label{fig:eff}
\end{figure*}

The importance of the combination of chemical and kinematical data can also be shown in Fig.~\ref{fig:eff}, in which we plot the probability distributions of these three subsamples. For each subsample, there are three kinds of probabilities, with  $P_{h\alpha}^\alpha$ for only using the [$\alpha$/M] data,  $P_{h\alpha}^{\boldsymbol{v}}$ for only using the velocities, and $P_{h\alpha}^{\alpha \boldsymbol{v}}$ for simultaneously using both of them. Respectively, the effectiveness indexes $E$ are calculated through Eq.~\ref{equ:eff}, and denoted as $E^\alpha $, $E^{\boldsymbol{v}}$ and $E^{\alpha \boldsymbol{v}}$. Comparing with these $E$ values, we can find that, the [$\alpha$/M] is the most effective variable to separate the \hamp-\lamp\ mixture (in the middle panel with the largest value of $E^\alpha$), while the kinematic data are very effective to separate the halo and \hamp\ components (the left panel). In all, when both [$\alpha$/M] and velocity data are involved, the effectiveness of the separation are increased, more or less.  \\

We can also calculate and compare the $E$ values for the whole high- and low-$\alpha$ sequences. The corresponding values are:  $E^\alpha = 0.88$, $E^{\boldsymbol{v}} = 0.67$ and $E^{\alpha \boldsymbol{v}} = 0.93$. It shows the improvement on the combination of chemical and kinematical data, and also demonstrate the high effectiveness of the membership determination of the current sample. \\

%----

\section{Comparison and Discussion}\label{sec:discussion}

Based on the chemical and kinematical data and the GMM method, the MW disk is divided into four components with distinctive chemical properties, high-$\alpha$ or low-$\alpha$, metal-poor or metal-rich, named as \hamp, \hamr, \lamp\ and \lamr\ disks respectively. Here we describe their properties individually, discuss the correlations among them and further analyze the implications for the formation scenario of the MW disk. The MW halo is also a separated component characterized by its chemical abundance and kinematics, but will not be discussed in this paper.

\subsection{Chemical Abundance}\label{subsec:abundance}

\begin{description}
  \item[\hamp\ disk] It is the major part of the high-$\alpha$ sequence, covering the [M/H] range from -1.2 to -0.1 dex, with monotonically decreased  [$\alpha$/M] from 0.29 to 0.14 dex. This significant [$\alpha$/M]-[M/H] anti-correlation could be reproduced by the chemical evolution model of a star forming process. Along this [$\alpha$/M]-[M/H] trend, the [$\alpha$/M] dispersion $\sigma_{[\alpha/{\rm M}]}$ is about 0.05 dex, varies from 0.035 to 0.074 dex. \\

  \item[\hamr\ disk] It extends the high-$\alpha$ sequence towards the most metal-rich point [M/H]$\sim 0.4$ dex, with [$\alpha$/M] decreasing from 0.10 to $\sim$ 0.02 dex.  In the transition area, the \hamp\ and \hamr\ disks seem to have breaks between them for the average values of number density and their distributions in the [$\alpha$/M]-[M/H] plane. For the main part of this disk, $\sigma_{[\alpha/{\rm M}]} \lesssim 0.04$ dex, which is slightly smaller than the \hamp\ disk.  \\

  \item[\lamp\ disk] It covers the [M/H] from -0.8 to -0.1 dex. This population distinct from the \hamp\ disk with a clear offset of [$\alpha$/M] $\sim$ -0.15 dex, while has the similar trend of the [$\alpha$/M]-[M/H] correlation, with [$\alpha$/M] decreases from 0.10 to 0.04 dex. The $\sigma_{[\alpha/{\rm M}]}$ is generally smaller than the high-$\alpha$ sequence, with the values around 0.025 dex. However, the number density contours in the upper-left panel of Fig.~\ref{fig:FittingResults} show a definite overlap of these two metal-poor disks. That is because they are not only determined by chemical abundance, but also by involving star's kinematical characteristics. Therefore, we declare that the dispersion of these two populations should be more reliable.\\

  \item[\lamr\ disk] Unlike the other three disk components, the \lamr\ disk appears almost constant mean values of [$\alpha$/M]$\sim 0.016$ dex across the [M/H] range from -0.1 to 0.4 dex, with similar dispersions $\sigma_{[\alpha/{\rm M]}} \sim 0.020$ dex, which is the smallest value among the four components. Although the number densities of \lamr\ disk may have strong overlaps with the \hamr\ disk, it is possible to distinguish them via the mean [$\alpha$/M] values, dispersions, and the [$\alpha$/M]-[M/H] trends.

\end{description}

In brief, these four disk components show different distribution features in the [$\alpha$/M]-[M/H] plane. The dispersions of the [$\alpha$/M] are vary among them, but generally agree with other works (e.g. $\sigma_{[\alpha/{\rm Fe}]} \sim$ 0.04 dex in \citealt{2021MNRAS.508.5903V}).  Moreover, since the observational error of [$\alpha$/M] is not too large, with a typical value $\sim 0.01$ dex for stars with [M/M] $>$ -1.0 dex, one can suppose that the [$\alpha$/M] dispersions are mainly due to the intrinsic distributions. The largest [$\alpha$/M] dispersion of the \hamp\ disk implies that it has been under the most violent star formation process. \\

\subsection{Kinematics}\label{subsec:kinematics}

The common kinematic features of four disk components are their average values ($\sim 0$ km s$^{-1}$) of radial velocity $V_R$, vertical velocity $V_Z$, and the symmetrical distributions of these two velocities. Obviously, these disks distinguish from each others through the rotational velocity $V_\phi$ and the velocity dispersions of all three velocities ($\sigma_{V_R}, \sigma_{V_\phi}, \sigma_{V_Z}$), as we itemize below.\\

\begin{description}
       \item[\hamp\ disk] The most significant feature of this disk is the  $V_\phi$-[M/H] correlation, with $V_\phi$ increase with [M/H] from 134 to 196 km\,s$^{-1}$. A liner fitting arrives a positive slope of  d$V_{\phi}$/d[M/H]= 49 km\,s$^{-1}$ dex$^{-1}$. This slope value is compatible with the earlier observational or simulative studies (\citealt{2011ApJ...738..187L,2013A&A...554A..44A,2018MNRAS.473..867K} ). On the other hand, the velocity dispersions, including $\sigma_{V_R}$, $\sigma_{V_\phi}$ and $\sigma_{V_Z}$, all monotonously decrease with [M/H] from 82 to 51 km\,s$^{-1}$,  76 to 35 km\,s$^{-1}$ and  71 to 31 km\,s$^{-1}$, respectively. \\ 
       
       \item[\hamr\ disk] Its $V_\phi \approx 200$ km s$^{-1}$, roughly keeps in constant within the metal-rich range. It is a little bit slower than the typical rotational speed of the MW. Velocity dispersions are all having small variations, with the average values of  $\sigma_{V_R} \sim$45 km\,s$^{-1}$, $\sigma_{V_\phi}\sim$32 km\,s$^{-1}$ and $\sigma_{V_Z}\sim$27 km\,s$^{-1}$. \\ 

       \item[\lamp\ disk] This disk generally obeys the typical rotation of the MW, with $V_\phi \sim$220 km\,s$^{-1}$ across its whole [M/H] coverage from -0.8 to -0.1 dex. A very small gradient with [M/H] could be found, which means the rotation of more metal-rich stars is a few km s$^{-1}$ slower. The slope is about -7 km\,s$^{-1}$dex$^{-1}$. It is significantly shallower than those slopes of \cite{2011ApJ...738..187L} whose subsample of thin disk is G-dwarfs, and \cite{2014A&A...567A...5R} whose sample of thin disk is FGK-type stars. The velocity dispersions $\sigma_{V_R} \sim 33$  km\,s$^{-1}$ and $\sigma_{V_\phi} \sim 22$ km\,s$^{-1}$. While the $\sigma_{V_Z}$ shows a gradient with [M/H] significantly, decreases from 28 to 15 km\,s$^{-1}$, which means the metal-poor part  (usually consists of older stars) of this disk has larger $V_Z$ dispersions. \\

       \item[\lamr\ disk] Its $V_\phi$ follows the small [M/H] gradients of the \lamp\ disk , and further decreases to 210 km s$^{-1}$. The velocity dispersions are approximately  remained constants with $\sigma_{V_R} \sim 35$ km\,s$^{-1}$, $\sigma_{V_\phi} \sim 22$ km\,s$^{-1}$, $\sigma_{V_Z} \sim 16$ km\,s$^{-1}$. It is the dynamically coldest disk.
\end{description}

In summary, the two metal-rich disks and the \lamp\ disk follow the rotation of the MW. The \hamp\ disk is slowly rotating, while the velocity increases with the metallicity. Based on the comparison of velocity dispersions, we can conclude that the \hamp\ is the hottest disk, while the \lamr\ is the coldest disk. The other two components could be regarded as warm disks but still are rotation-dominated populations. \\

It is interesting that there are breaks between the two high-$\alpha$ disks, in the trends of  $V_\phi$-[M/H] and $\sigma_V$-[M/H] for all three velocity dispersions. For these four kinematical parameters,  the values increase/decrease with [M/H] for the \hamp\ disk, but keep in constants for the \hamr\ disk. These kinds of breaks imply the change of star formation procedure. In contrast, the kinematics of the \lamp\ and \lamr\ disks are quite similar, and there are no significant breaks between them, except the $\sigma_{V_Z}$ which is obviously larger of the  \lamp\ disk than that of the \lamr\ disk.\\

Additionally, one may find that,  for all disks and in all [M/H] bins except the $\sigma_{V_Z}$ of the most metal-poor  part of the \lamp\ disk, we have $\sigma_{V_R} > \sigma_{V_\phi} > \sigma_{V_Z}$.  It follows the general relations of the whole MW disk from previous works \citep{2003A&A...410..527B,2010ApJ...716....1B}.\\

Finally, it is worth noting that, for the low-$\alpha$ sequence, at the [M/H]$\sim -0.8$ dex, there is an extremely sudden break of $V_\phi$ and dispersions of all three velocities. This is a very clear separation of the \lamp\ disk and the halo, though there is a continuity in the chemical abundance distributions.\\

\subsection{Spatial distribution}\label{subsec:spatial}

\begin{figure*}[htbp]
	\centering
	\includegraphics[width=1.00\textwidth]{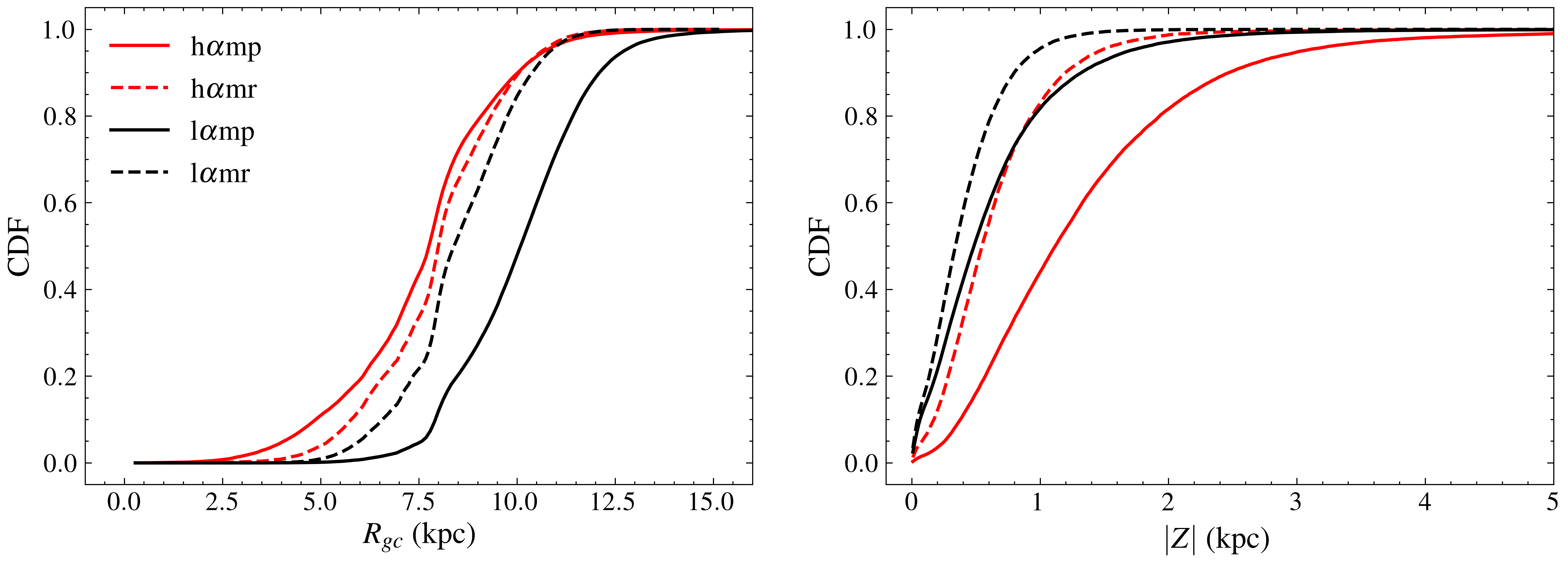}
	\caption{Left panel: The cumulative $R_{\rm gc}$ distributions for the \hamp, \hamr, \lamp\ and \lamr\ disks. Right panel: The cumulative $|$Z$|$ distributions for the same four disks.}
	\label{fig:cdfRZ}
\end{figure*}

Our sample is limited and shaped by the APOGEE survey, and is not volume completed. So it is difficult to accurately model the spatial distribution of individual disk components, e.g., to measure their scale-length or scale-height. Nevertheless, we can still compare their relative distributions. For this purpose, we adopt the cumulative distribution function (CDF) of $R_{\rm gc}$ and $|Z|$. Fig.~\ref{fig:cdfRZ} shows these two CDFs of the four disk components individually. Here we accumulate the membership probability of each star that belongs to the specific disk component.\\

In the left panel of Fig.~\ref{fig:cdfRZ}, we can find that, the \hamp\ disk is the innermost component with its members mainly distribute from $R_{\rm gc} =$ 3.5 to 10 kpc. While the \lamp\ disk stars mainly distribute farther than 7.5 kpc from the Galactic center, and extend to $\sim$15 kpc. The other two metal-rich disks, \hamr\ and \lamr\ are somewhat in between of them, whereas more close to the \hamp\ disk. Therefore, we can roughly define the \hamp, \hamr, and \lamr\ as three inner disks, while the \lamp\ as an outer disk. Moreover, along the chemical abundance evolution trace, i.e., the increase of [M/H] or the decrease of [$\alpha$/M], the sizes of the three inner disks, from \hamp\ to \hamr\ and then to \lamr,  enlarge in sequence. \\

The right panel of Fig.~\ref{fig:cdfRZ} shows the pattern of three thin disks verse one thick disk. The \lamr\ disk is the thinnest one, with its sample stars concentrated within $\pm 1$ kpc from the Galactic plane. The thickest disk is the \hamp\, which extends its member stars to higher than 4 kpc. The other two intermediate disks look quite similar in the $|Z|$ distributions. However, since the \lamp\ disk has a much larger $R_{\rm gc}$ distribution, it should be even thinner than the \hamr\ disk. Interestingly, considering the three inner disks, \hamp, \hamr\ and \lamr, they are not only enlarged, but also getting thinner in sequence.   \\

Combined with the $R_{\rm gc}$ and $|Z|$ distributions, we can conclude that the \hamp\ disk is the only real thick (hot) disk with the inner most distribution towards the Galactic center. The two low-$\alpha$ disks, either the outer or the inner ones, are thin (cold) disks. Without considering the \lamp\ disk, the other three inner disks follow the inside-out and upside-down formation scenario (\citealt{2006ApJ...639..126B,2013ApJ...773...43B,2015ApJ...804L...9M}).  These results are also consistent with recent analysis by \cite{2021A&A...647A..73S} using the APOGEE DR16. \\

\subsection{Age distribution}\label{subsec:age-distribution}

\begin{figure*}[htbp]
	\centering
	\includegraphics[width=1.00\textwidth]{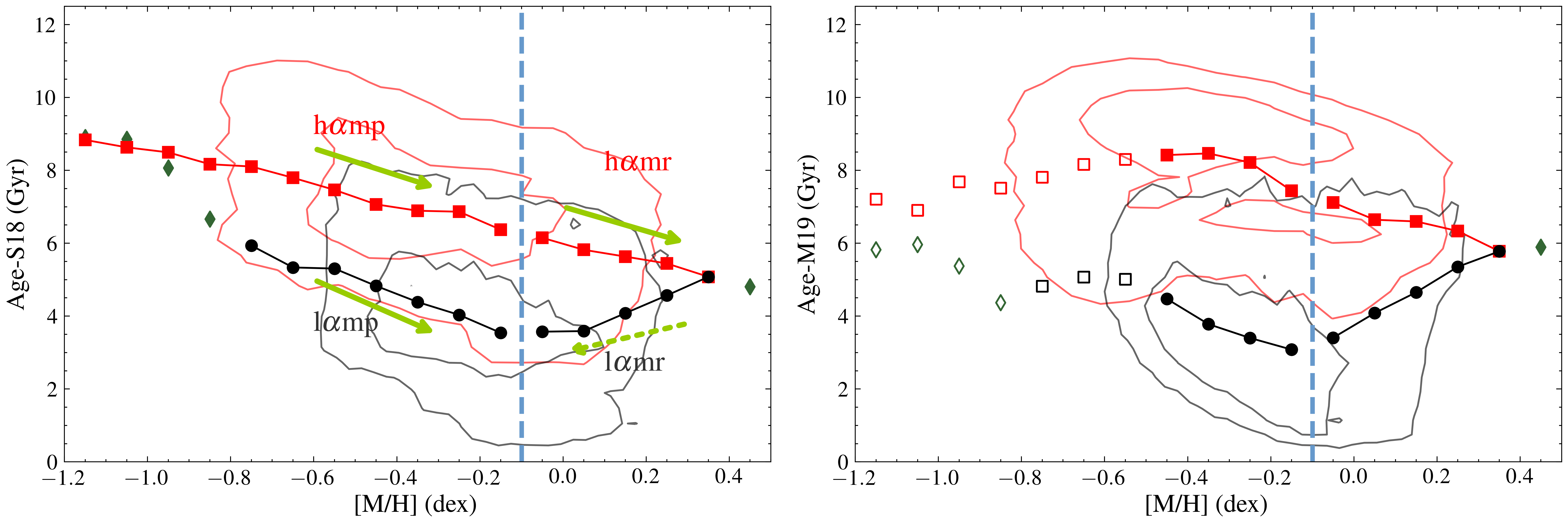}
	\caption{The age-metallicity distributions. Left and right panels are for Age-S18 \citep{2018MNRAS.481.4093S} and Age-M19 \citep{2019MNRAS.489..176M} respectively.  The red and black contours represent the 1 and 2$\sigma$ probability-weighted number densities of high- and low-$\alpha$ disks, respectively. The symbols are probability-weighted mean ages of each metallicity bin. Green arrows in the left panel hint the evolution trace of four disks.  As mentioned by \cite{2019MNRAS.489..176M}, the Age-M19 of stars with [Fe/H] $< -0.5 $ dex are less reliable, hence we use hollow symbols for [M/H] $< -0.5 $ dex. }
	\label{fig:age}
\end{figure*}

For most of our sample stars, we have the age measurements from two sources (see Sec.~\ref{subsec:age} for details), Age-S18 and Age-M19. Although both of them have large uncertainties for individual stars, it is surely possible for us to statistically determine the typical age of a subsample within a given [M/H] bin, and then explore the age features of four disks. It is worth noting that, in the GMM, we only use the chemical and kinematical data, so the age could be regarded as an independent variable in the discussion.\\

Fig.~\ref{fig:age} shows the probability-weighted number densities of high- and low-$\alpha$ sequences as contours in the Age-[M/H] plane.  Age-S18 and Age-M19, are plotted in two panels separately. The symbols show the average age of each disk component in each [M/H] bin (also listed in table~\ref{tab:result}). Since the Age-M19 of stars with [Fe/H] $< -0.5 $ dex are less reliable than the others \citep{2019MNRAS.489..176M}, we use hollow symbols for [M/H] poorer than -0.5 dex, and neglect them in following analysis. \\

With the current stellar age measurements, either Age-S18 or Age-M19, large dispersions ($\sim$1.5 Gyr) of both high- and low-$\alpha$ sequences are present. However, there is an overall trend that metal-rich stars are noticeably younger than metal-poor stars. It is also clear that there is a time lag between two sequences, with the high-$\alpha$ populations roughly 2$\sim$3 Gyrs older than the low-$\alpha$ ones.\\

Three disk components, the \hamp, \hamr, and \lamp, have similar correlations between age and [M/H], which indicates the more metal-rich stars are averagely younger than the more metal-poor stars. These trends obey the normal abundance enhancement procedure of stellar formation for a population. The age distribution of the \hamp\ disk is in rough agreement with that of the canonical thick disk stars previously discussed by many works (e.g., \citealt{2007ApJ...663L..13B,2018A&A...619A.125A,2021A&A...645A..85M,2021A&A...654A..13L}). \\

On the contrary, and more importantly, this kind of correlation of the \lamr\ disk is inverse, with younger stars tending to be more metal-poor than older stars. This phenomenon is clearly shown for both Age-S18 and Age-M19. The abnormal Age-[M/H] trend might be explained as there is a sufficient gas rejuvenation in the later forming stage of this disk. It will cause the enhancement of the star-forming activity leading to much more stars in the \lamr\ disk than in the \hamr\ disk, which is exactly the case we can show in their fractions and number density distributions (see in Fig.~\ref{fig:FittingResults}). It also causes an interesting phenomenon that the youngest population of our MW is not the most metal-rich but has [M/H]$\sim$-0.1, which is a little bit poorer than the Sun. \\

Alternatively, if we compare the stars of the low-$\alpha$ sequence along the timeline, one may notice that the older stars have wide, or probably dichotomous metallicity distribution, while the younger stars' distribution is gradually getting narrower and converges to [M/H]$\sim$-0.1 dex. This phenomenon is even more obvious for the Age-M19 distribution. It implies that in the early stage of these two low-$\alpha$ disks, their stars formed separately in outer and inner regions by irrelevant gas with significantly different metalicities. Then, with ongoing gas enrich and mixture processes, the later formed stars tend to have similar metalicities. \\

From another point of view, the three inner disks seem to conjoint in a consequent formation process, whereas the outer disk, \lamp, is more likely to occupy an independent formation process. In the outer region, it enhances the disk itself, and also brings metal-poor gas into the  inner region to mix with metal-rich gas and then leading to the abnormal Age-[M/H] relationship of the \lamr\ disk. \\

\subsection{Differences among disks}\label{subsec:diskDiff}

Among the four disk components, there are conjoint components, two high-$\alpha$ disks, two low-$\alpha$ disks, and two heavily mixed metal-rich disks. Actually, there are pieces of evidence to identify the independence of these stellar populations, as we summarize below.  \\

\begin{description}
  \item[\hamp\ $vs$ \hamr] In the [$\alpha$/M]-[M/H] plane (see red contours of the top-left panel of Fig.~\ref{fig:FittingResults}), there is a sudden decrease of [$\alpha$/M] from \hamp\ to \hamr, and also a possible saddle point of the number density, at [M/H]$\simeq -0.1$ and [$\alpha$/M]$\simeq 0.14$. The most significant difference between these two disks is the change of the $V_\phi$ and the break of the $[V_\phi]$-[M/H] correlation. Also, the velocity dispersions $\sigma_V$ of the \hamp\ disk are all larger than the \hamr\ disk. Another significant difference is that the $|Z|$ dispersion of \hamp\ stars are much higher than those of the \hamr\ disk.\\

  \item[\lamp\ $vs$ \lamr] There is no break in the number density distribution of these two disks. However, there are breaks in the [$\alpha$/M]-[M/H] and $\sigma_{V_Z}$-[M/H] correlations. One of the obvious differences is in their $R_{\rm gc}$ distributions, where the \lamp\ stars mostly located in the outer regions, which may lead to the well known radial gradient of metallicity. The most important difference is shown in the Age-[M/H] trends, where the \lamr\ disk follows an inverse abundance evolution trace with younger stars having poorer metallicity. This is the dominant reason to distinguish the conventional thin disk into two parts.   \\

  \item[\hamr\ $vs$ \lamr] These two disks are directly and rigorously determined by the GMM.  So the smaller BIC value for a two components fit ($K=2$) is the first evidence of the existence of two metal-rich disks. Besides the difference of the [$\alpha$/M] distributions, the \hamr\ disk also has lower $V_{\phi}$ and larger dispersions for all three velocities than the \lamr\ disk. Of cause, the most important difference between them is their inverse Age-[M/H] trends. Generally, the \hamr\ stars are older than the \lamr\ stars at the same metallicity.  This characteristic has been mentioned by \cite{2011A&A...535A.107K} and \cite{2021A&A...654A..13L}.\\
 
\end{description}

It is commonly agreed that there is a transition zone between the thick disk and the metal-rich part of the thin disk. This stellar family, also named as the \hamr\ disk were already reported and investigated by many previous works with different star samples (\citealt{2011A&A...535L..11A,2013MNRAS.430..836N,2013A&A...554A..44A,2018MNRAS.473..867K,2021ApJ...912..106Y}). It is still an open question whether this population only represents the metal-rich tail of the thick (\hamp) disk, or the \hamr\ stars might have originated from the inner Galactic bulge and have migrated towards the solar neighborhood. Unfortunately, previous works usually employed a simple cut of higher $\alpha$-enhancement stars at the metal-rich region. While in this work, we use the GMM to divest the \hamr\ disk from the metal-rich part of the thin (low-$\alpha$) disk, so it can rigorously derive all features of this disk. This rigorous statistical solution also benefits the characterizing of the remaining \lamr\ component.\\

In summary, the \hamr\ and \lamr\ are two particular disks that we have divested, identified, and quantitatively described rather than in previous works.  \\

\subsection{Implication of the disk star formation scenario}\label{subsec:scenario}

Our results suggest a formation scenario of the MW disks involving two pathways of star formation and chemical evolution, which generally supports the two-infall picture of Milky Way (e.g. \citealt{1997ApJ...477..765C,2020A&A...635A..58S}).　The first pathway is from the \hamp\ disk to the \hamr\ disk, and then to the \lamr\ disk. It occurs the initial starburst with its consequent process. The canonical thick (\hamp) disk forms earlier from the highly turbulent gas with a broad vertical spread. This episode is very short and confirms the results of other works  (\citealt{1997ApJ...477..765C,2020A&A...635A..58S}). After the rapid star formation is quenched, it transfers to a long-term secular evolution phase with lower SFR, leading to also inner but kinematically cold disk components. In the later stage of this star-forming episode, the remaining gas was further cooled down but continually rejuvenated with lower [M/H] gas from the second gas accretion, resulting in the gradually decrease of the average stellar metallicity. In the view of kinematics, these three components satisfy a process of going from hot to cold and accelerating rotation.  \\

The second pathway is from the \lamp\ disk to the younger part of the \lamr\ disk. It may be triggered by another accretion of a huge amount of low [M/H] gas about 3 Gyr later than the first starburst and occurs the formation of the outer thin disk (the \lamp\ disk), and the significant time lag confirms the results of previous works (\citealt{2019A&A...623A..60S,2020A&A...635A..58S,2021A&A...647A..73S}). Then, the metal-poor or somewhat enriched gas continually rejuvenated the most metal-rich gas of the inner part, and gradually dominate the \lamr\ disk's recent star formation. Comparing with the correlation slopes of Age-[M/H] of the two metal-poor disks, we can find the slope of \lamp\ disk is steeper than the \hamp\ disk. That means this second star formation process may not be as violent as the earlier one. The discrepancy of the $\alpha$-enhancement and the velocity dispersions between these two disks also support this conclusion, since the $\alpha$-enhancement is much more efficiency for the shorter time scale star formation.  \\  

This two pathways scenario also leads to a complicated formation process of the \lamr\ disk and reminds us that the metal-rich region ([M/H]$>$-0.1) is far away from the detailed investigation. Although the separation of \hamr\ and \lamr\ disks may not be the only deconstruction, it provides a novel view in understanding the origin of the metal-rich stars.  \\

This overall MW disk formation scenario reminds several interesting MAP points. One is at [M/H]$\simeq -0.1$ and [$\alpha$/M]$\simeq 0.14$, which is the breakpoint of two high-$\alpha$ disks. According to the age measurements used in this paper, this phase-transfer happened at $\sim$6.5 (Age-S18) or $\sim$7.5 (Age-M19) Gyrs ago. Another break point is between the \lamp\ disk and halo components, at [M/H]$\simeq -0.8$ and [$\alpha$/M]$\simeq 0.1$. This point indicates the beginning of the second gas accretion happened about 6.5 Gyrs ago. Besides, the most metal-rich population of the MW disk is at [M/H]$\simeq 0.45$ and [$\alpha$/M]$\simeq 0.0$, which is the turn-off point showing the gas rejuvenation making effects about 5.0 (Age-S18) or 6.0 (Age-M19) Gyrs ago, and is roughly 1.5 Gyrs after the quench of the first starburst and the beginning of the second gas infall. Moreover, the youngest MAP of the star sample of this work is at [M/H]$\simeq -0.1$ and [$\alpha$/M] $\simeq 0.0$, which is a little bit more metal-poor than the Sun and represents the chemical abundance of the remained and well mixed cool gas. We noted that, although the current age measurements for a large sample of stars may still have large uncertainties or unneglectable bias, the relative order and the time lag of different populations are reliable. \\ 

Our results are also consistent with the well-known inside-out and upside-down MW disk formation picture (\citealt{2006ApJ...639..126B,2013ApJ...773...43B,2015ApJ...804L...9M}). The overall scenario is caused by the two major gas infall processes mainly represented by two metal-poor disks. The early formed \hamp\ disk is significantly smaller and thicker (hotter) than the second accretion occurred \lamp\ (outer) disk. Besides this large scale picture, if we only consider the three inner disks of the first episode, we can also discover an inside-out and upside-down behavior. Along the formation sequence, from \hamp\ to \hamr, and then to \lamr, these inner disks are sequentially enlarged in size and getting thinner in thickness. \\　

We have to mention that the MW dynamical processes, e.g., the radial migration \citep{2009MNRAS.399.1145S,2009MNRAS.396..203S} or the heating process \citep{2009ApJ...707L...1B} by the accretion of dwarf galaxies (or sub-halos), may significantly modulate the kinematics and spatial distributions of different disk components formed in different epochs. Therefore, a more detailed kinematical analysis, including the information of the stellar orbit properties, energy, and angular momentum, is required. We may extensively discuss these issues in a separate paper.

\section{Conclusion}\label{sec:conclusion}

This work aims to quantify the chemical and kinematical properties of the Galactic disk components to understand their structure formation and evolution. We use a large sample of 119,558 giants from APOGEE DR17 that cross-match Gaia EDR3. By employing the GMM method for sub samples of metallicity bins, we obtain the distribution parameters (mean and dispersion values) of [$\alpha$/M], $V_R$, $V_\phi$, and $V_Z$ as functions of [M/H] (table~\ref{tab:result} and Fig.~\ref{fig:FittingResults}). So the dichotomous distribution feature is then identified and quantified, not only in the $\alpha$-enhancement but also in the velocity spaces. However, we still use high- and low-$\alpha$ to name the distinct sequences which can well characterize them. As a by-product, we calculate the membership probabilities belonging to each sequence for individual sample stars (table~\ref{tab:Pmemb}), and the membership determination effectiveness is also evaluated (Fig.~\ref{fig:binFit} and ~\ref{fig:eff}). \\

These two sequences show significant overlaps in the distributions on [$\alpha$/M]-[M/H] and $V$-[M/H] planes which confirm the necessity of the combination of multiple variables in separating these two sequences. Obviously, properties vary along the sequences, so we can determine four disk components with distinctive chemical properties and name them as \hamp, \hamr, \lamp, and \lamr\ disks, respectively. Compared with previous works \citep{2005A&A...438..139S,2006MNRAS.367.1181B,2011A&A...535L..11A,2014A&A...562A..71B}, the most improvement of this classification is that we divest the \hamr\ disk from the metal-rich part with a rigorous statistical method and distinguish the \lamr\ disk from the canonical thin disk. It makes the \lamr\ disk a particular and well-defined component. Additionally, two specific MAPs are figured out as the breakpoints between the two high-$\alpha$ disks and between the \lamp\ disk and  halo populations.\\

Besides the difference in chemical abundance, four disk components show their difference in the properties of kinematics, spatial distributions (Fig.~\ref{fig:cdfRZ}), $V$-[M/H] correlations (Fig.~\ref{fig:FittingResults}), or Age-[M/H] correlations (Fig.~\ref{fig:age}). The varieties between these disks generally support the two-infall formation scenario of the MW disk (e.g., \citealt{1997ApJ...477..765C, 2017MNRAS.472.3637G,2019A&A...623A..60S,2020MNRAS.497.2371L}). The first infall of turbulent gas quickly forms the hot thick disk and consequently changes to the secular phase with a lower star formation rate to form the \hamr\ and \lamr\ disks \citep{2020MNRAS.497.2371L,2021A&A...647A..73S}. The second major gas accretion is more temperate to form the \lamp\ disk that dominates the outer Galactic disk region and rejuvenate the inner disk part to influence the secular formation process  of the \lamr\ disk. It results in a more metal-poor young population. \\

We have discovered that the inside-out and upside-down disk forming scenario is hierarchical. The \hamp\ disk, which is the canonical thick disk that dominates the first gas infall is significantly smaller and thicker than the \lamp\ formed with the later gas accretion (\citealt{2006ApJ...639..126B,2013ApJ...773...43B,2015ApJ...804L...9M}). Meanwhile, the three inner disks, \hamp, \hamr, and \lamr, also show the evidence of getting extender and thinner successively. \\

The most interesting phenomenon we have revealed is the inverse Age-[M/H] trend of the \lamr\ disk, which means that younger stars are more metal-poor than older stars of this disk component. It violates the metal increasing trend of a spatially enclosed star-forming area and indicates that, in the inner Galactic disk region, the rejuvenate gas gradually takes the initiative of the later star formation. \\

Finally, from the perspective of the entire Galactic disk, the recently formed stars convergence to [M/H]$\sim$-0.1 dex. It demonstrates that the gas of the second accretion is sufficiently mixed with the remains of the first gas infall.   \\

% \acknowledgments
We thank Ruixiang Chang, Maosheng Xiang for the helpful discussions. This work is supported by National Key R\&D Program of China No. 2019YFA0405501, the National Natural Science Foundation of China (NSFC) under grant U2031139, and the science research grants from the China Manned Space Project with NO. CMS-CSST-2021-A08. This work has made use of data from the European Space Agency (ESA) mission Gaia (https://www.cosmos.esa.int/gaia), processed by the Gaia Data Processing and Analysis Consortium (DPAC; https://www.cosmos.esa.int/web/gaia/dpac/consortium). Funding for the DPAC has been provided by national institutions, in particular the institutions participating in the Gaia Multilateral Agreement. 

The stellar parameters, abundances, RVs from APOGEE DR17 were derived
from the allStar files available at https://www.sdss.org/dr17/. Funding for the Sloan Digital Sky Survey IV has been provided by the Alfred P. Sloan Foundation, the U.S. Department of Energy Office of Science, and the Participating Institutions. \\
%==============================================================
\software{Astropy \citep{2013A&A...558A..33A}, Numpy \citep{2011CSE....13b..22V}, Scipy \citep{2007CSE.....9c..10O}, Matplotlib \citep{2007CSE.....9...90H}, \texttt{emcee} \citep{2013PASP..125..306F}.}

\newpage
	
	\bibliography{ref}{}
	\bibliographystyle{aasjournal}

\end{CJK*}	
\end{document}